\documentstyle[12pt,epsf]{article} 
\begin{document} 
\thispagestyle{empty} 
\noindent 
{\bf FZR-161\\ 
December 1996\\ 
Preprint\\[2cm]} 
\begin{center} 
{\large \bf 
Relativistic Description of Exclusive Deuteron\\ 
Break-up Reactions}\\[1cm] 
{\sc 
L.P. Kaptari$^{a,b}$, 
B. K\"ampfer$^{a,c},$ 
S.M. Dorkin$^d$, 
S.S. Semikh$^b$}\\[1cm] 
$^a$Research Center Rossendorf, Institute for Nuclear and Hadron Physics,\\ 
PF 510119, 01314 Dresden, Germany\\[1mm] 
$^b$Bogoliubov Laboratory of Theoretical Physics, JINR Dubna,\\ 
P.O. Box 79, Moscow, Russia \\[1mm] 
$^c$Institute for Theoretical Physics, Technical University Dresden,\\ 
01062 Dresden, Germany \\[1mm] 
$^d$ Far-Eastern State University, Vladivostok, Russia 
\end{center} 
%
%
\vskip 1cm 
\begin{abstract} 
The exclusive deuteron break-up reaction is analyzed 
within a covariant approach based on the Bethe-Salpeter 
equation with realistic meson-exchange interaction. 
Relativistic effects 
in the cross section, tensor analyzing power 
and polarization transfer are 
investigated in explicit form. Results of numerical calculations 
are presented for kinematical conditions in forthcoming 
p + D reactions at COSY. 
\end{abstract} 
 
\vspace*{1cm} 
 
\noindent 
key words: exclusive deuteron break-up, Bethe-Salpeter equation, 
polarization observables\\[3mm] 
PACS number(s): 21.45+v, 21.10.Ky, 21.60.-n 
\newpage 
\setcounter{page}{1} 
\noindent 
{\bf 1. Introduction:} 
Break-up reactions of 
deuterons by protons receive presently a renewed interest 
\cite{cosy,experiment,dpelastic,proposal}. 
The investigations of break-up processes are motivated by the hope 
to extract directly the deuteron wave function from experimental data, 
supposed the mechanism described by the impulse approximation dominates 
at moderate values of intrinsic momenta. 
Pioneering experimental studies of elastic 
backward and inclusive  D + p reactions have been performed 
in Dubna and Saclay \cite{experiment,dpelastic,proposal}. 
It is found, indeed, that the impulse approximation holds 
in a large interval of momenta of detected protons, 
except for a broad shoulder 
at momentum of the outgoing proton (measured in the deuteron's rest frame) 
of about 0.3 GeV/c. 
Probably small corrections to the impulse approximation are sufficient to 
account for this shoulder. 
In the same experiments the polarization observables of 
the deuteron, such as the tensor analyzing power $T_{20}$ and 
polarization transfer $\kappa$ are measured. 
These quantities are more sensitive to the 
reaction mechanism, and a combined analysis of them may 
provide information about $S$ and $D$ components of the 
deuteron wave function separately. 
However, as in the unpolarized case, the data on $T_{20}$ and $\kappa$ 
exhibit systematic deviations from theoretical predictions 
\cite{break,lykasov} in the same momentum region of the outgoing proton. 
A kinematical analysis of the invariant missing mass shows 
that this region corresponds to $\Delta$ excitations, 
so that additional corrections should supplement the calculations 
based on the impulse approximation. From this it becomes clear 
that an interpretation of the 
inclusive data in terms of a deuteron one-body momentum density becomes 
ambiguous. Furthermore, the typical energies in these reactions seem already 
too high for a non-relativistic approach. 
It is necessary, therefore, to describe these processes in a 
covariant formalism. 
 
The measurement of  proton-deuteron break-up reactions 
with polarized particles in an exclusive experimental setup is planned 
at the cooler synchrotron COSY in J\"ulich \cite{cosy}. 
The experiment will detect 
the scattered fast proton in the forward direction in the spectrometer 
ANKE in coincidence with the slow backward-emitted proton. 
The kinematical conditions are chosen in such a way that the missing mass 
of the undetected particles 
is exactly equal to the neutron mass. 
Hence, this experiment offers a unique possibility of using polarized deuterium 
targets in combination with a polarized proton beam for extended 
studies of exclusive deuteron break-up processes. The 
coincidence measurement allows one to exclude kinematically 
particle production processes and, consequently, 
to constrain the reaction mechanism. 
Two important aspects in these experiments 
ought to be stressed here: 
(i) The possibility of excluding $\Delta$ excitation processes 
will allow for understanding better the role of the meson 
degrees of freedom and to clarify the origin of the above mentioned 
shoulder in the inclusive processes. 
(ii) A comparison with theoretical predictions 
will provide a good test of the validity of the spectator mechanism. 
 
Previous relativistically invariant investigations \cite{break,keister} 
of one-nucleon exchange diagrams for inclusive and elastic 
D + p scattering 
are based on numeric solutions of the Bethe-Salpeter (BS) 
equation with a realistic interaction. Several authors \cite{gross} 
studied relativistic effects in the deuteron by considering 
the  D $\to$ NN vertex within approximations to the 
exact BS equation and analyzed so electromagnetic and 
elastic hadron scattering off the deuteron. Up to now, however, a consistent 
investigation of polarization phenomena in  exclusive 
processes with explicit identification of relativistic effects 
is still lacking. 
 
In the present work we perform such a covariant analysis 
of the exclusive proton - deuteron break-up reaction. 
Fully covariant expressions 
for the cross section, tensor analyzing power  $T_{20}$ 
and transferred polarization $\kappa$ 
are obtained within the BS formalism. 
The contribution of pure relativistic corrections is separated. 
Results of numerical calculations, utilizing the 
recently obtained numerical solution of the BS equation 
with a realistic interaction kernel, 
are presented for kinematical conditions available at COSY. 
%
%
 
\noindent 
{\bf 2. The spectator mechanism:} 
We consider the above described exclusive break-up reaction of the type 
p + D $\to$ p$_1$(0$^o$) + p$_2$(180$^o$) + n$_3$(0$^o$). 
Within the spectator mechanism approach 
this reaction is presented by the Feynman diagram shown in fig.~1a, where 
the upper and lower vertices factorize and, consequently, 
they can be computed separately. 
According to the above kinematical conditions one may consider 
the fast forward proton to be produced by elastic 
scattering of the beam proton off one nucleon in the deuteron. 
In fig.~1b this  part of the diagram is depicted and 
the corresponding cross section reads 
\begin{equation} 
d^6 \sigma^{NN}= 
\frac{1}{2\sqrt{\lambda(p_p,p_n)}} 
\frac{\delta^4 (p_p+p_n-p_1-p_3)}{(2\pi)^2} \, 
|T_{NN}(p_p,p_n)|^2  \, 
\frac{d^3p_1}{2E_1} \, 
\frac{d^3p_3}{2E_3}, 
\label{dp1} 
\end{equation} 
where $\lambda(p_p,p_n) = (p_p p_n)^2-m_p^2 m_n^2$ and 
the  invariant nucleon-nucleon (NN) amplitude $T_{NN}$ is expressed 
via the truncated vertex functions $\Gamma_{NN}$ 
and projectors $\Lambda_{\alpha\alpha'}$ as follows 
\begin{eqnarray} 
&& 
|T_{NN}|^2 = \frac{1}{4} \, \Lambda_{\alpha\alpha'}(p_p) \, 
\Lambda_{\beta\beta'}(p_n) \, 
{\cal O}_{\alpha\beta;\alpha'\beta'},\nonumber\\ 
&& 
{\cal O}_{\alpha\beta;\alpha'\beta'} =\sum\limits_{spins} 
\left ( u(p_3,s_3) \bar \Gamma_{NN} 
u(p_1,s_1) \right )_{\alpha'\beta'} 
\left ( \bar u(p_1,s_1) \Gamma_{NN} \bar u(p_3,s_3) \right )_{\alpha\beta}, 
\label{dp2} 
\end{eqnarray} 
where $u$ denote nucleon Dirac spinors and $\alpha, \beta$ are Dirac indices 
(with summation over twofold indices) 
and the operator (\ref{dp2}) is related to 
the $T$ matrix of nucleon-nucleon scattering. 
To calculate the latter one 
in a fully covariant manner one needs an analysis of the NN 
vertices within  an appropriate relativistic formalism, e.g., the BS 
approach. For the elastic scattering 
the NN amplitude is thoroughly investigated 
in ref.~\cite{{bselastic}}. 
In our case the operator ${\cal O}$ 
determines the differential elastic cross section $d\sigma/dt$ 
at low transferred 
momentum $t=(p_p-p_1)^2$, 
and the contribution of the spin 
part in eq.~(\ref{dp2}) may be neglected. Then the operator 
${\cal O}_{\alpha\beta;\alpha'\beta'}$ takes the simplest possible form 
$ 
{\cal O}_{\alpha\beta;\alpha'\beta'}\approx 
\delta_{\alpha\alpha'}\delta_{\beta\beta'} |A_{NN}|^2, 
$ 
where $ A_{NN}$ is the scalar part of the NN amplitude. 
Then, with $m$ as nucleon mass, 
\begin{equation} 
d\sigma^{NN} = \frac{m^2}{4\pi\lambda(p_p,p_n)}|A_{NN}|^2\, dt. 
\label{dp4} 
\end{equation} 
Now we are in the position to calculate the main contribution to the 
process described by the diagram in fig.~1a. 
Applying the Mandelstam method the invariant 
cross section for the break-up of a polarized 
deuteron with spin projection $M$ can be written as 
\begin{eqnarray} 
&& 
2E_2\frac{d^4\sigma}{dt \, d^3p_2}= 
\frac{1}{(2\pi)^3} 
\frac{\sqrt{\lambda(p_p,p_n)}}{\sqrt{\lambda(p_p,p_D)}} \, 
\frac{d\sigma^{NN}}{dt} \, 
\frac{(p_2^2-m^2)}{2m} \, 
\mbox{Tr} \left ( \bar \Psi_M^D(p)\hat I \, \Psi_M^D(p) (\hat p_2-m)\right), 
\label{dp5} 
\end{eqnarray} 
where 
$p=(p_n-p_2)/2$, and 
$\hat I$ stands either for a unity matrix in case when the outgoing 
proton $p_2$ is not polarized or for 
$(1+\gamma_5\hat s_2)/2$ otherwise ($\hat s_2$ is 
the contracted polarization four vector). 
$\Psi_M^D(p)$ denotes the charge conjugated 
BS amplitudes introduced in ref.~\cite{khanna}; numerical solutions are 
reported in refs.~\cite{solution,parametrization}. 
 
Eq.~(\ref{dp5}) is a rather formal result 
and an interpretation of different contributions is straightened  in the 
present form. A few subtle aspects are to be mentioned:\\ 
(i) The detected proton is on mass-shell so that 
$p_2^2-m^2=0$, and the cross section seems to vanish. 
However, the BS amplitude itself is 
singular when one nucleon is on mass-shell. 
To get a finite result the corresponding expressions 
should be evaluated analytically.\\ 
(ii) The numerical solution of the BS equation is obtained 
in the Euclidean space-time, where the time component $p_0$ of 
the relative momentum $p$ is purely imaginary. 
In  processes under consideration $p_0$ is fixed and real. 
Hence, one needs either a numerical procedure for an analytical 
continuation of the amplitudes to the real relative 
energy axis (cf. \cite{keister}) or another recipe \cite{anatomia} 
for using  our numerical solutions for this case .\\ 
(iii) In solving the BS equation we expand the 
amplitude $\Psi_M^D$ on the complete set of 
the Dirac matrices and obtain eight partial amplitudes 
for  $\Psi_M^D$ \cite{khanna,quad}. 
However, it is known \cite{keister,gross} that in such cases, 
where one nucleon is on mass-shell, only four partial amplitudes 
contribute to the deuteron observables. From eq.~(\ref{dp5}) 
it is not clear which amplitudes play the most important role 
in the process. 
 
To tackle the above problems it is convenient 
to transform our representation of the partial amplitudes to the so-called 
$\rho$ spin classification \cite{quad,cubis}. 
In this representation one usually adopts the spectroscopic notation for the 
partial amplitudes $^{(2S+1)} L_J^{\rho_1\rho_2}$. 
(For the explicit form of the unitary transformation matrix 
between these two representations cf.~\cite{quad}.) 
In this notation it becomes immediately clear that 
in  processes with one nucleon  (say the second one) on mass-shell 
the relevant contribution to  the cross section comes 
only from four amplitudes with positive second $\rho$ spin index, 
i.e., $^3S^{++}_1,\, ^3D^{++}_1,\, ^1P^{-+}_1 $ and $^3P^{-+}_1$ 
for which we use the short hand notation 
$\Psi_S,\, \Psi_D,\, \Psi_{P_1}$ and $\Psi_{P_3}$, respectively. 
%
%
 
\noindent 
{\bf 3. Relativistic corrections:} 
The trace in eq. (\ref{dp5}) is 
evaluated by an algebraic formula manipulation code which 
delivers the contribution 
to the unpolarized cross section 
\begin{eqnarray} 
&&\frac{(p_2^2-m^2)}{2m} 
\frac{1}{3} 
\sum\limits_M Tr  \left [ \bar \Psi_M^D(p)\hat I \Psi_M^D(p) (\hat p_2-m)\right]= 
\frac{(p_2^2-m^2)}{8\pi E_2} 
(2E_2+2p_0-M_D)\nonumber\\ 
&& 
\times 
\left\{ 
\frac{1}{2} 
\left [ -  \Psi_S^2(p_0,|{\bf p}|)- \Psi_D^2(p_0,|{\bf p}|) 
+ \Psi_{P_1}^2(p_0,|{\bf p}|)+ \Psi_{P_3}^2(p_0,|{\bf p}|)\right ] \right .\label{dp6}\\ 
&& 
\quad 
+ 
\frac{\sqrt{3}}{3} 
\frac{|{\bf p}|}{m} 
\left [ \Psi_S(p_0,|{\bf p}| ) \left (- \Psi_{P_1}(p_0,|{\bf p}|) 
+\sqrt{2} \Psi_{P_3}(p_0,|{\bf p}|) \right )\right.\nonumber\\ 
&& 
\quad 
- 
\left.\left. 
\Psi_D(p_0,|{\bf p}| ) \left (\sqrt{2} \Psi_{P_1}(p_0,|{\bf p}|)+ 
\Psi_{P_3}(p_0,|{\bf p}|) 
\right) \right] 
\phantom{\frac12}\right\}, 
\nonumber 
\end{eqnarray} 
where  another zero due to $ 2E_2+2p_0-M_D = 0$ appears explicitly 
in the cross section ($M_D$ is the deuteron mass). 
To handle these two zeros and the singularities 
in the amplitudes $\Psi^D_M$ it is convenient to introduce instead of 
$\Psi^D_M$ the corresponding 
partial vertices $G(p_0,|{\bf p}|)$ 
that have no poles when one particle is on mass shell. For an explicit 
relation between partial amplitudes and partial vertices 
we refer the interested reader to ref.~\cite{quad}, where 
the dependence of $S$ and $D$ wave 
vertices upon the relative energy is shown to be 
smooth, contrary to the amplitudes which 
display a strong dependence on $p_0$. Therefore, in our  calculations, 
we can replace, at moderate values of $p_0$, 
the $S$ and $D$ vertices by their values at $p_0=0$ with good accuracy. 
The $P$ vertices can be expand into Taylor series 
about $p_0=0$ up to a desired order in $p_0/m$. Then the corresponding 
derivatives can be computed numerically 
along the imaginary axis since they are 
analytical functions of $p_0$ \cite{anatomia}. 
After replacing the amplitudes by vertices, 
the zeros and singularities  cancel at $p_{20}\to E_2$, 
and the result is finite. Finally, to cast 
our formulae more familiar form, known from non-relativistic 
calculations, we introduce the notion of BS wave functions 
\cite{keister,gross,quad} 
\begin{equation} 
U(|{\bf p}|) = {\cal F}\frac{G_S(0,|{\bf p}|)}{2E_2-M_D}, 
\quad 
W(|{\bf p}|) = {\cal F}\frac{G_D(0,|{\bf p}|)}{2E_2-M_D}, 
\quad 
V_{P_{1(3)}}(|{\bf p}|) = {\cal F}\frac{G_{P_{1(3)}}(0,|{\bf p}|)}{M_D}, 
\label{vertex} 
\end{equation} 
where ${\cal F} = 1/4\pi\sqrt{2M_D}$. 
With these definitions one gets\\ 
(i) the unpolarized differential cross section: 
\begin{eqnarray} 
E_2\frac{d^4\sigma}{dt d^3p_2} &\equiv& 
\frac{M_D}{2\pi^2}\frac{\sqrt{\lambda(p_p,p_n)}} {\sqrt{\lambda(p_p,p_D)}} 
\frac{d\sigma^{NN}}{dt}\cdot \frac{1}{3}\sum\limits_M D_M(|{\bf p}|) 
\nonumber\\ 
\hspace*{1cm} 
&=& 
\frac{M_D}{2\pi^2} 
\frac{\sqrt{\lambda(p_p,p_n)}} 
{\sqrt{\lambda(p_p,p_D)}} 
\frac{d\sigma^{NN}}{dt}\left\{ \phantom{\frac{1}{2}} 
\hspace*{-3mm} 
\left[ U^2(|{\bf p}|)+W^2(|{\bf p}|) - V_{P_1}^2(|{\bf p}|) 
- V_{P_3}^2(|{\bf p}|) \right ] 
\right. \label{dp8} \\ 
&& 
\hspace*{-3cm} 
+ 
\left. 
\frac{2\sqrt{3}}{3}\frac{|{\bf p}|}{m} 
\left[U(|{\bf p}|) \left ( 
- V_{P_1}(|{\bf p}|)+\sqrt{2} V_{P_3}(|{\bf p}|)\right ) 
-W(|{\bf p}|) \left ( 
\sqrt{2} V_{P_1}(|{\bf p}|)+ V_{P_3}(|{\bf p}|)\right ) 
\right ]\, \,\right\}, 
\label{dp9} 
\end{eqnarray} 
(ii) the tensor analyzing power $T_{20}$: 
\begin{eqnarray} 
\left( \frac{1}{3}\sum\limits_M D_M(|{\bf p}|)\right ) 
\sqrt{2}T_{20}  &=& 
\hspace*{-2mm} 
\left [- W^2(|{\bf p}|) - 2\sqrt{2} U(|{\bf p}|) W(|{\bf p}|) 
+2 V_{P_1}^2(|{\bf p}|) - V_{P_3}^2(|{\bf p}|) 
\right ] \label{dp10}\\ 
&& 
\hspace*{-5.7cm} 
+ 
\frac{2\sqrt{3}}{3}\frac{|{\bf p}|}{m} 
\left[2U(|{\bf p}|) \left ( 
V_{P_1}(|{\bf p}|)+ V_{P_3}(|{\bf p}|)/\sqrt{2}\right ) 
+W(|{\bf p}|) \left ( 
2\sqrt{2} V_{P_1}(|{\bf p}|)- V_{P_3}(|{\bf p}|)\right ) 
\right ],\label{dp11} 
\end{eqnarray} 
(iii) the polarization transfer $\kappa$: 
\begin{eqnarray} 
\left ( \frac{1}{3}\sum\limits_M D_M(|{\bf p}|)\right )\kappa 
& = & 
\left [ U^2(|{\bf p}|) - W^2(|{\bf p}|) 
+\frac{\sqrt{2}}{2} U(|{\bf p}|) W(|{\bf p}|)\right ] \label{dp12}\\ 
&& 
\hspace*{-5cm} 
+ 
\sqrt{\frac{3}{2}} 
\frac{|{\bf p}|}{m} 
\left[U(|{\bf p}|) \left ( 
-\sqrt{2} V_{P_1}(|{\bf p}|)+ V_{P_3}(|{\bf p}|)\right ) 
+W(|{\bf p}|) \left ( 
 V_{P_1}(|{\bf p}|)+ \sqrt{2}V_{P_3}(|{\bf p}|)\right ) 
\right ]. 
\label{dp13} 
\end{eqnarray} 
For short hand notation we introduce the deuteron structure factor 
$D_M(|{\bf p}|)$ with a definition which follows from 
eqs.~(\ref{dp8},~\ref{dp9}). 
 
A numerical analysis of solutions of the BS equation in terms 
of amplitudes within the $\rho$ spin basis shows \cite{quad} that the BS 
wave functions $U(|{\bf p}|)$ and  $W(|{\bf p}|)$ are the dominant ones 
and coincide to a large extent with 
the corresponding non-relativistic wave functions found as solutions of the 
Schr\"odinger equation with one-boson-exchange potential. 
The remaining two functions  $V_{P_{1,3}}(|{\bf p}|)$ are a 
few orders of magnitude smaller, and for the considered processes 
with momentum of the backward proton $|{\bf p}_2|=|{\bf p}| \le$ 0.51 GeV/c 
the diagonal terms in  $V_{P_{1,3}}(|{\bf p}|)$ are negligible. 
Therefore, eqs.~(\ref{dp8},~\ref{dp10},~\ref{dp12}) 
are identified as the main  contributions 
to the corresponding observables  and they 
might be compared with their non-relativistic analogues. 
The interference terms (\ref{dp9},~\ref{dp11},~\ref{dp13}) 
posses contributions from  negative states and are 
proportional to $|{\bf p}|/m$. Due to 
their pure relativistic origin we refer to them as 
relativistic corrections in the deuteron break-up reactions. 
Note that, when disregarding the relativistic corrections and 
equating the wave functions $U(|{\bf p}|)$ and  $W(|{\bf p}|)$ 
to their non-relativistic analogues, our formulae 
(\ref{dp8} - \ref{dp13}) exactly recover the non-relativistic 
expressions for $\sigma$, $T_{20}$, $\kappa$ 
computed within the spectator mechanism 
\cite{lykasov}. 
%
%
 
\noindent 
{\bf 4. Results and discussions:} 
In figs.~2 - 4 we present the results of our calculations 
by exploiting the numerical solutions 
\cite{solution,parametrization,quad} 
of the BS equation 
with a realistic interaction kernel with 
$\pi,\omega,\rho,\sigma,\eta,\delta$ exchange 
(parameters as in table 1 in ref.~\cite{solution}). 
The unpolarized cross section computed by 
eqs.~(\ref{dp8},~\ref{dp9}) 
is displayed in fig.~2, 
where the flux factor has been put to 
unity. This flux factor reflects only 
the dependence of the cross section on the beam energy within the 
spectator mechanism approach. By disregarding it one obtains the 
energy independent part of the cross section. 
Within the spectator mechanism, both $T_{20}$ and $\kappa$ 
do not depend on the initial energy. 
For the elastic neutron - proton scattering 
we use a fit of data \cite{data}. 
The dashed curves in figs.~2 - 4 
depict the contribution of only  positive 
waves according to eqs.~(\ref{dp8}, \ref{dp10}, \ref{dp12}), while 
the short-dashed curves are the relativistic corrections 
according to eqs.~(\ref{dp9},~\ref{dp11},~\ref{dp13}) 
(in fig. 2 we display the modulus of the corrections, because of a 
sign change), and the 
full lines are their sums. For completeness we also present the results of 
non-relativistic calculations with the Bonn potential wave functions shown 
as dotted curves. 
 
It is seen in figs.~2 - 4 that the relativistic corrections are negligible 
small in a wide range of momenta $|{\bf p}_2|$ 
and become significant only at 
$|{\bf p}_2| > 0.6$ GeV/c. From this we conclude that 
at kinematical conditions as envisaged in COSY experiments \cite{cosy} 
the relativistic corrections may be safely neglected since they are much 
smaller than the expected experimental errors. 
At $E_p\sim$ 3 GeV one has 
$|{\bf p}_2|_{max}\approx$ 0.51 GeV/c. 
(Note that within the spectator mechanism 
the maximum value of $|{\bf p}_2|$ is 
restricted to $|{\bf p}_2|_{max}\sim$ 0.8 GeV/c.) 
Hence, in the proposed experiments one may investigate in great detail 
different aspects of the reaction  mechanism, or the contribution 
of non-nucleonic degrees of freedom like meson-exchange currents and 
the role of $\Delta$ isobars, without bothering about relativistic effects. 
Note that our expressions for the observables in exclusive break-up 
processes are similar to those for the inclusive ones 
\cite{break,lykasov}. 
There exist an essential advantage in the proposed exclusive 
experiments: in inclusive processes the detected 
shoulder~\cite{experiment} in the cross section at 
$|{\bf p}_2|\approx$ 0.3 GeV/c cannot be explained within 
the spectator mechanism. The origin of the discrepancy 
is believed \cite{lykasov} to root in the contribution of meson production in 
the NN vertex. In the exclusive processes these 
contribution may be separated kinematically and conclusions 
about this problem can be settled. 
 
The relativistic corrections eqs.~(\ref{dp9}, \ref{dp11}, \ref{dp13}) 
are governed by 
negative $P$ states in the deuteron. This can be considered 
as a hint that admixtures of $P$ waves within the BS approach 
are related to relativistic corrections by 
taking into account meson-exchange currents and N$\bar{\mbox{N}}$ 
pair production diagrams \cite{mec} in the non-relativistic picture. 
To establish a correspondence between our results and  the 
mentioned non-relativistic calculations we estimate 
the contribution of the  relativistic corrections by 
computing the $P$ wave vertices in the so-called 
``one-iteration approximation''. 
The gist of this approximation is as follows \cite{karmanov}: in solving the 
BS equation by an iteration procedure one puts as  zeroth 
iteration the exact solution of the Schr\"odinger equation for 
$S$ and $D$ vertices and zero for other waves; then the $P$ 
vertices are found by one iteration of the BS equation. 
Our experience in solving numerically the BS equation shows that 
it converges rapidly for relatively small momenta $< 1$ GeV/c. That means 
when utilizing  the exact non-relativistic solutions, 
after one iteration the resulting $P$ waves are not too far from the reality. 
 
Skipping cumbersome algebraic manipulations the result for 
the function $V$ in eq.~(\ref{vertex}) with a 
BS kernel with pseudo-scalar one-boson exchange reads as follows 
\begin{equation} 
V_{P_{1,3}}(|{\bf p}|) = -g_pi^2 
\frac{2\sqrt{3}}{M_DE_2} 
\int\limits_0^\infty dr \, \frac{{\rm e}^{-\mu r}}{r} \, (1+\mu r) \, 
{\rm j_1}(|{\bf p}| r) 
\left [ 
N_u \, U(r) + N_w \, W(r)\right ],\label{uwr} 
\end{equation} 
where $U(r)$ and $W(r)$ are the non-relativistic deuteron wave functions 
in the coordinate representation, and $g_\pi^2 \approx$ 14.5 
is the pion-nucleon coupling constant; 
$N_u = 1 \, (\sqrt{2})$ and $N_w = \sqrt{2} \, (-1)$ for 
$P_1 \, (P_3)$ waves. 
Introducing this result into the expression 
for the cross section eq.~(\ref{dp8}) the relativistic corrections is 
\begin{eqnarray} 
E_2\frac{d^4 \sigma^{r.c.}}{dt d^3p_2} &=& 
\frac{\sqrt{\lambda(p_p,p_n)}} {\sqrt{\lambda(p_p,p_D)}} 
\frac{d\sigma^{NN}}{dt} 
\left( 
\frac{8}{\pi^2}\frac{{ v}}{c} 
\frac{g^2_\pi}{4m} \right. 
\int dr 
\frac{{\rm e}^{-\mu r}}{r} \, (1+\mu r) \, 
{\rm j_1}(|{\bf p}| r) \nonumber\\ 
&\times& 
\left. 
\left\{ 
U(|{\bf p}|)[-U(r)+2\sqrt{2}W(r)] +  W(|{\bf p}|)[2\sqrt{2} U(r) 
+W(r)]\right\} 
\phantom{\frac12} \right), 
\label{dp14} 
\end{eqnarray} 
which is similar to expressions obtained in non-relativistic evaluations 
of the so-called ``catastrophic'' and pair production   diagrams 
in electro-disintegration of 
the deuteron \cite{foldy}. In eq.~(\ref{dp14}) 
$v/c$ is the velocity of the detected slow proton, 
and the quantity in the large parenthesis may be interpreted 
as effective number of N$\bar{\mbox{N}}$ pairs in the deuteron contributing 
to break-up reactions 
(details will be presented elsewhere \cite{inpreparation}). 
From this it becomes clear that 
generic relativistic calculations, even in impulse approximation, 
contain already some specific meson-exchange diagrams, i.e., 
pair production currents, and one should pay attention 
on the problem of double counting when computing relativistic corrections 
beyond the spectator mechanism. 
 
\noindent 
{\bf 5. Summary:} 
In summary, we present for the first time 
an explicit analysis of relativistic effects 
in exclusive deuteron break-up reactions within the Bethe-Salpeter 
formalism with realistic interaction kernel. 
Numerical estimates of relativistic effects in the cross section, 
tensor analyzing power and polarization transfer 
at kinematical conditions of forthcoming COSY experiments are performed. 
Relativistic corrections under these conditions are identified and 
found negligible. 
It is shown that the planned experiments can discover mainly 
effects related to processes beyond the impulse approximation.\\ 
{\bf Acknowledgments:} 
Useful discussions with M. Beyer, S. Bondarenko, H. M\"uller and 
A.Yu. Umnikov are gratefully acknowledged. 
One of the authors (L.P.K.) would like 
to thank for the warm hospitality in the Research Center Rossendorf. 
 
\newpage 
\newpage 
\centerline{\bf Figure captions} 
 
\vspace*{1cm} 
 
\noindent 
Fig.~1: 
Feynman graphs for the exclusive reaction 
p + D = p$_1$(0$^o$) + p$_2$(180$^o$) + n$_3$(0$^o$) (a) and 
for the elementary NN vertex (b).\\[1cm] 
Fig.~2: 
The spin averaged differential cross section 
$E \, d \sigma / dt\,d^3p_2$ (with flux factor put to 1) 
for the exclusive proton - deuteron break-up reaction. 
The meaning of the curves are explained in the text. 
\\[1cm] 
Fig.~3: 
The deuteron tensor analyzing power $T_{20}$ for the exclusive 
proton - deuteron break-up reaction. 
Notation as in fig.~2.\\[1cm] 
Fig.~4: 
The polarization transfer $\kappa$ for the exclusive 
proton - deuteron break-up reaction. 
Notation as in fig.~2. 
 
\newpage
\epsfbox{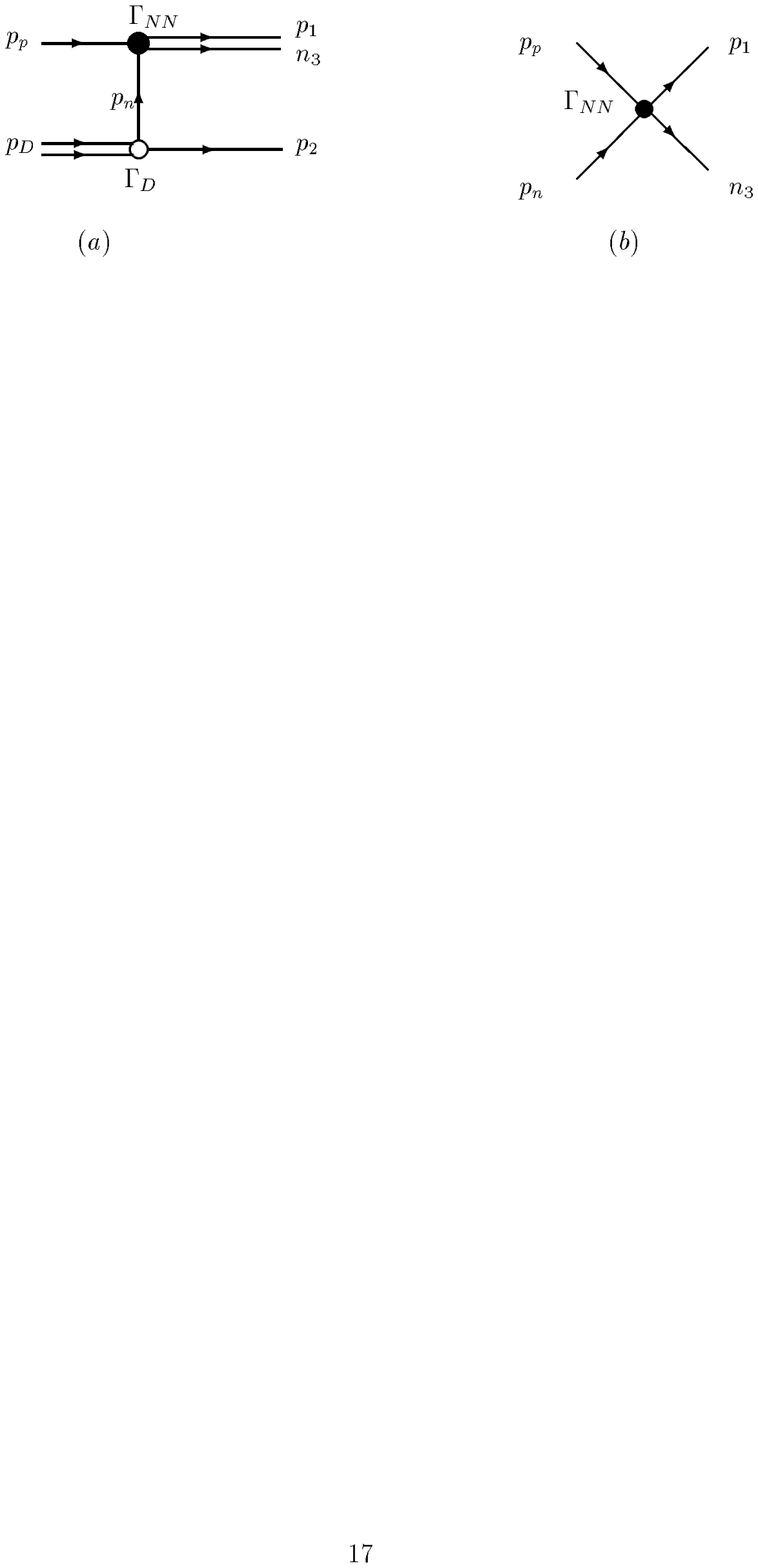}\vspace*{-15cm}
Fig.~1: 
Feynman graphs for the exclusive reaction 
p + D = p$_1$(0$^o$) + p$_2$(180$^o$) + n$_3$(0$^o$) (a) and 
for the elementary NN vertex (b).
\newpage 
\epsfxsize 6in
\vspace*{-3cm} \hspace*{2cm} \epsfbox{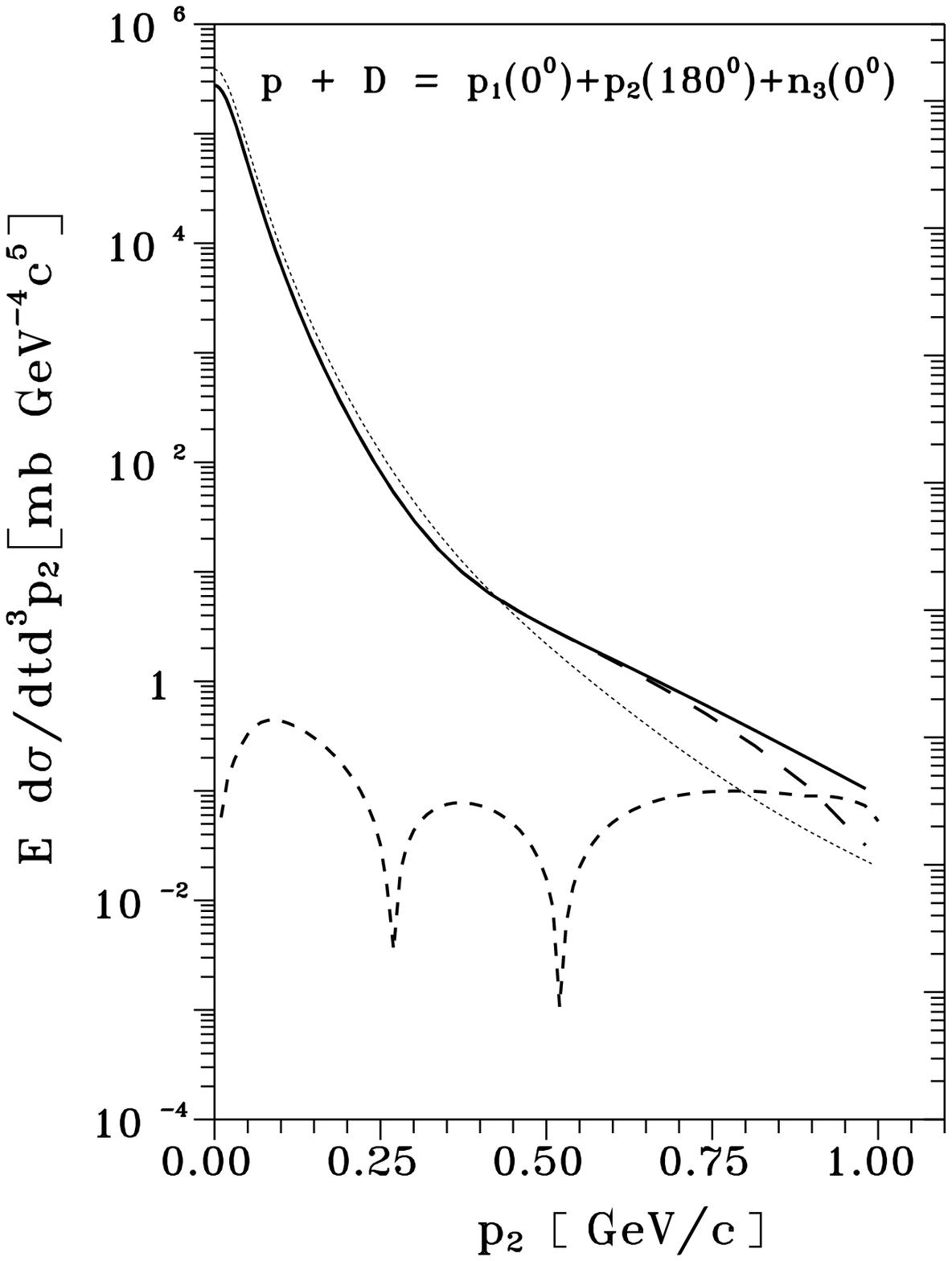}

 Fig. 2: 
The spin averaged differential cross section 
$E \, d \sigma / dt\,d^3p_2$ (with flux factor put to 1) 
for the exclusive proton - deuteron break-up reaction. 
The meaning of the curves are explained in the text.

\newpage 
\epsfxsize 7in
\vspace*{-3cm}\epsfbox{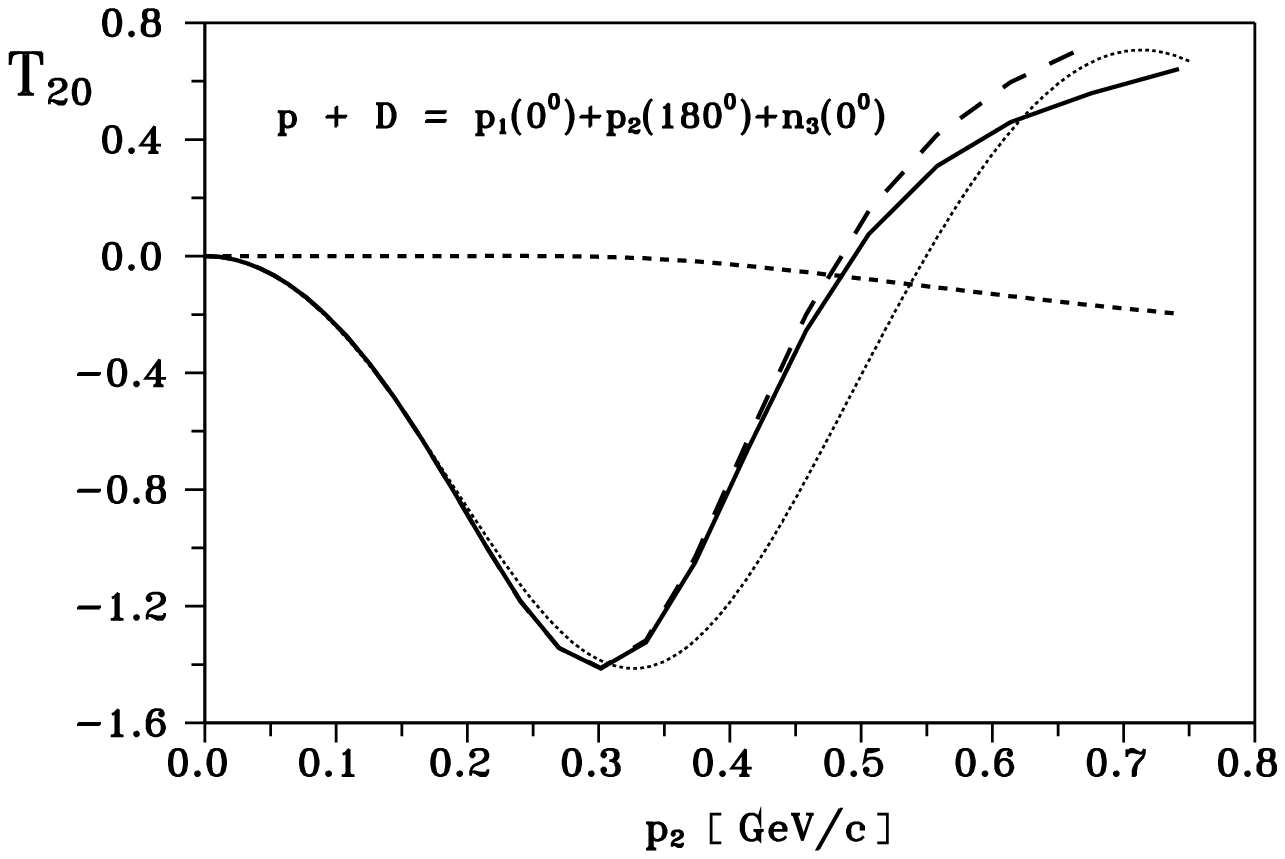} 
 
Fig.~3: 
The deuteron tensor analyzing power $T_{20}$ for the exclusive 
proton - deuteron break-up reaction. 
Notation as in fig.~2. 
\newpage 
\epsfxsize 7in
\vspace*{-3cm}
\epsfbox{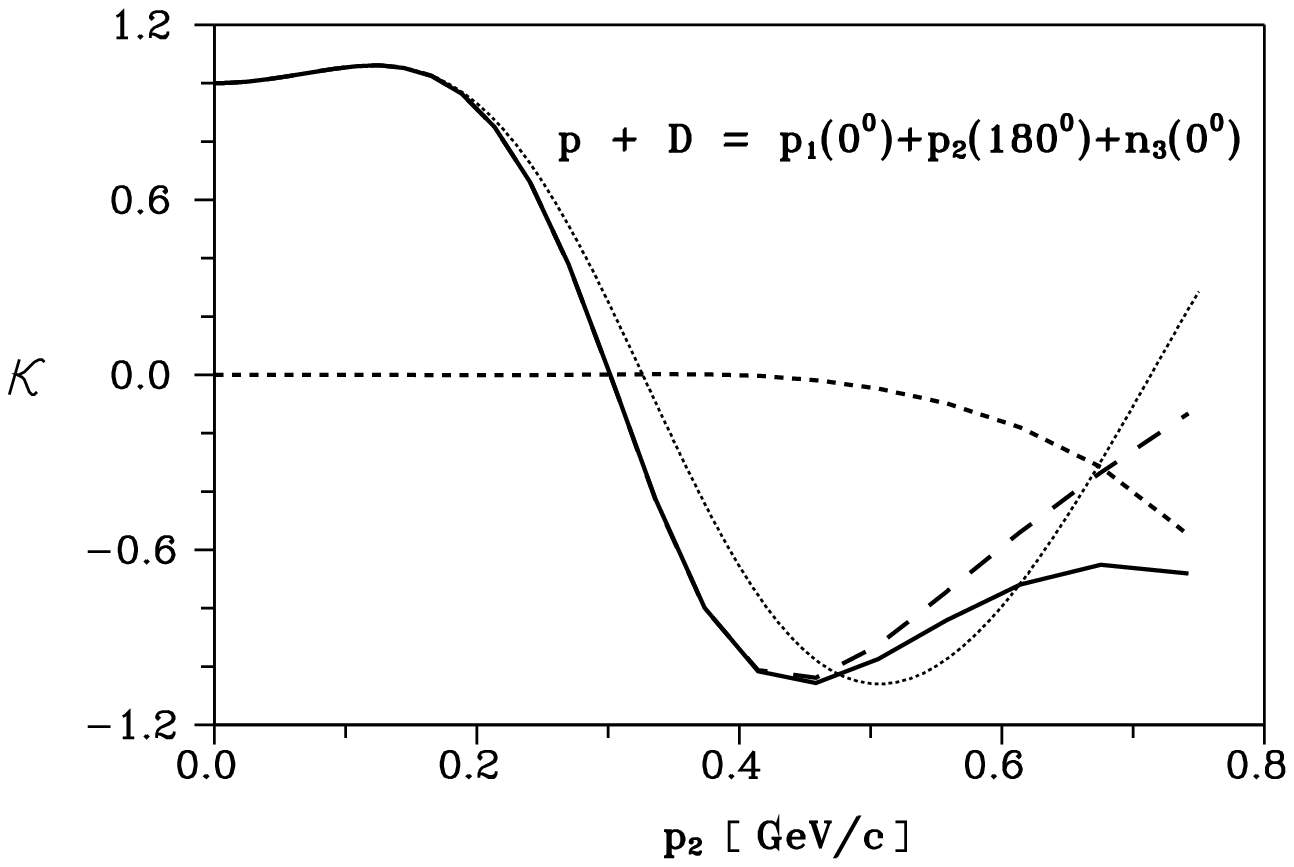} 
 
Fig.~4: 
The polarization transfer $\kappa$ for the exclusive 
proton - deuteron break-up reaction. 
Notation as in fig.~2. 
\newpage 

\end{document}